\documentclass[12pt]{article}
\usepackage{setspace}
\usepackage{epsfig}
\usepackage{amssymb}
\def\be{\begin{equation}}
\def\ee{\end{equation}}
\def\ba{\begin{array}}
\def\ea{\end{array}}
\def\beqn{\begin{eqnarray}}
\def\eeqn{\end{eqnarray}}

\def\bt{\begin{tabular}}
\def\et{\end{tabular}}
\def\bc{\begin{center}}
\def\ec{\end{center}}

\begin{document}

\title{Exploring the likelihood of CP violation in neutrino oscillations}

\author{Gulsheen Ahuja\\
\\
{\it Department of Physics, Centre of Advanced Study, P.U.,
Chandigarh, India.}\\ {\it Email: gulsheenahuja@yahoo.co.in}}

\maketitle

\begin{abstract}
In view of the latest T2K and MINOS observations regarding the
mixing angle $s_{13}$, we have explored the possibility of
existence of CP violation in the leptonic sector. Using hints from
the construction of the `$db$' unitarity triangle in the quark
sector, we have made an attempt to construct the `$\nu_1.\nu_3$'
leptonic unitarity triangle, suggesting a good possibility of
having non zero CP violation.
\end{abstract}

\section{Introduction}
The recent T2K \cite{t2k} and MINOS \cite{minosnew} observations
regarding the mixing angle $s_{13}$ has given a big impetus to the
sharpening of the implications of the neutrino oscillations, in
particular the non zero value of angle $s_{13}$ implies the
possibility of existence of CP violation in the leptonic sector.
In the context of neutrino oscillation phenomenology, the last few
years have seen impressive advances in fixing the neutrino mass
and mixing parameters through solar \cite{solexp1}-\cite{solexp7},
atmospheric \cite{atmexp}, reactor (KamLAND \cite{kamland}, CHOOZ
\cite{chooz}) and accelerator (K2K \cite{k2k}, MINOS
\cite{minosnew,minos}) neutrino experiments. Adopting the three
neutrino framework, several authors
\cite{kamland,schwetztortolavalle}-\cite{foglinew} have presented
updated information regarding these parameters obtained by
carrying out detailed global analyses. In particular,
incorporating the above mentioned developments regarding the angle
$s_{13}$, Fogli {\it et al}. \cite{foglinew} have carried out a
global three neutrino oscillation analysis, yielding
\be
 \Delta m_{21}^{2} = 7.58^{+0.22}_{-0.26}\times
 10^{-5}~\rm{eV}^{2},~~~~
|\Delta m_{31}^{2}| = 2.35^{+0.12}_{-0.21}\times 10^{-3}~
\rm{eV}^{2},
 \label{solatmmass}\ee
\be
{\rm sin}^2\,\theta_{12}  =  0.312^{+0.017}_{-0.016},~~~
 {\rm sin}^2\,\theta_{23}  =  0.42^{+0.08}_{-0.03},~~~
 {\rm sin}^2\,\theta_{13}  =  0.025\pm 0.007.~~~\label{s12s23} \ee

In analogy with the quark mixing phenomenon, the above value of
$s_{13}$ which is not so `small' suggests likelihood of CP
violation in the leptonic sector. A comparison of the mixing
angles in the leptonic sector with those in the quark sector point
out that the CP violation could, in fact, be considerably large in
this case. This possibility, in turn, can have deep
phenomenological implications. As is well known, the two CP
violating Majorana phases do not play any role in the case of
neutrino oscillations, therefore any hint regarding the value of
Dirac-like CP violating phase in the leptonic sector $\delta_l$
will go a long way in the formulation of proposals on observation
of CP violation in the Long BaseLine (LBL) experiments
\cite{k2k,minos,opera}. In the absence of any hints from the data
regarding leptonic CP violation, keeping in mind the parallelism
between the neutrino mixing and the quark mixing, an analysis of
the quark mixing phenomena could provide some viable clues
regarding this issue in the leptonic sector.

It may be noted that in the context of fermion mixing phenomena,
the Pontecorvo-Maki-Nakagawa-Sakata (PMNS) \cite{pmns1,pmns2} and
the Cabibbo-Kobayashi-Maskawa (CKM) \cite{ckm1,ckm2} matrices have
similar parametric structure. Also, regarding the three mixing
angles corresponding to the quark and leptonic sector, it is
interesting to note that in both the cases the mixing angle
$s_{13}$ is smaller as compared to the other two. Taking note of
these similarities of features, in this paper, using the analogy
of the quark mixing case we have made an attempt to find the
possibility of the existence of CP violation in the leptonic
sector.

The detailed plan of the paper is as follows. In Section
(\ref{quark}), we examine the quark mixing case for obtaining
useful hints for the case of lepton mixing, studied in
Section(\ref{lepton}). Finally, in Section (\ref{summ}) we
summarize and conclude.

\section{Quark mixing case \label{quark}}
In this context, we first examine the case of quark mixing,
wherein the CKM matrix as well as the existence of CP violation
are well established. Parallel to the leptonic sector wherein only
the three mixing angles or correspondingly the magnitudes of the
three elements of the mixing matrix are known, one would like to
consider a similar situation in the quark sector and examine
whether one can deduce any viable information regarding the
existence of CP violation in the quark mixing phenomena. In this
context, several important features in the case of CKM paradigm
having implications for CP violation may be kept in mind. In
particular, one may note that the magnitudes of the CKM matrix
elements are rephasing invariant implying that these are
independent of the 36 parameterizations of the CKM matrix. Also
one has to keep in mind that in any parameterization the CP
violating phase of the CKM matrix can be related to the magnitudes
of its elements. Therefore, in case one can find the magnitudes of
the CKM matrix elements, even approximately, one may be able to
get some idea about the CP violation. In this context, the
Particle Data Group (PDG) representation \cite{pdgnew} of the CKM
matrix facilitates the approximate construction of the magnitudes
of the CKM matrix elements from the three well known mixing angles
or the elements $V_{us}$, $V_{cb}$ and $V_{ub}$, with one of the
angles being much smaller than the other two. The approximate
magnitudes of the CKM matrix elements then provide a way to
construct the unitarity triangles which can be used to determine
the Jarlskog's rephasing invariant parameter $J$, an important
parameter as all CP violating effects are proportional to it.
Further, with the knowledge of $J$, the CP violating phase
$\delta$ can be determined in the PDG representation. The same is
true in the leptonic sector also.

To this end, we first construct the approximate magnitudes of the
elements of the quark mixing matrix. For ready reference as well
as to facilitate discussion of results, the
Cabibbo-Kobayashi-Maskawa (CKM) matrix \cite{ckm1,ckm2} is defined
as
 \be \left( \ba {c} d~^{\prime} \\ s~^{\prime} \\ b~^{\prime} \ea \right)
  = \left( \ba{ccc} V_{ud} & V_{us} & V_{ub} \\ V_{cd} & V_{cs} &
  V_{cb} \\ V_{td} & V_{ts} & V_{tb} \ea \right)
 \left( \ba {c} d\\ s \\b \ea \right),  \label{qm}  \ee
which in the PDG representation \cite{pdgnew} involving angles
$\theta_{12}, \theta_{23}, \theta_{13}$ and the phase $\delta$ is
given as
   \be V_{CKM}=\left( \ba{ccl} c_{12} c_{13} & s_{12} c_{13} &
  s_{13}e^{-i \delta} \\ - s_{12} c_{23} - c_{12} s_{23}
  s_{13} e^{i \delta} & c_{12} c_{23} - s_{12} s_{23}
  s_{13} e^{i \delta} & s_{23} c_{13} \\ s_{12} s_{23} - c_{12}
  c_{23} s_{13} e^{i \delta} & - c_{12} s_{23} - s_{12} c_{23}
  s_{13} e^{i \delta} & c_{23} c_{13} \ea \right), \label{qmm} \ee
 with $c_{ij}= {\rm cos}~ \theta_{ij}$ and $s_{ij}= {\rm sin}~
\theta_{ij}$ for $i,j=1,2,3$.

Making use of the fact the mixing angle $s_{13}$ ($\equiv V_{ub}$)
is small in comparison to both $s_{12}$ ($\equiv V_{us}$) and
$s_{23}$ ( $\equiv V_{cb}$) as well as using their values given by
PDG 2010 \cite{pdgnew} and the above representation of the CKM
matrix, the approximate magnitudes of the CKM matrix elements come
out to be \be V_{CKM}= \left( \ba{ccc}
  0.97431\pm0.00021 & 0.2252\pm0.0009  &  0.00389\pm0.00044\\
  0.2250\pm0.0009  &  0.97351\pm0.00021  & 0.0406\pm0.0013\\
 0.00914\pm0.00029 &0.0396\pm0.0013 &0.999168\pm0.000053
 \ea \right). \label{ckmmag} \ee
The above matrix is in fairly good agreement with the one given
recently by PDG 2010 \cite{pdgnew}.

 \begin{figure}[tbp]
\centerline{\epsfysize=2.6in\epsffile{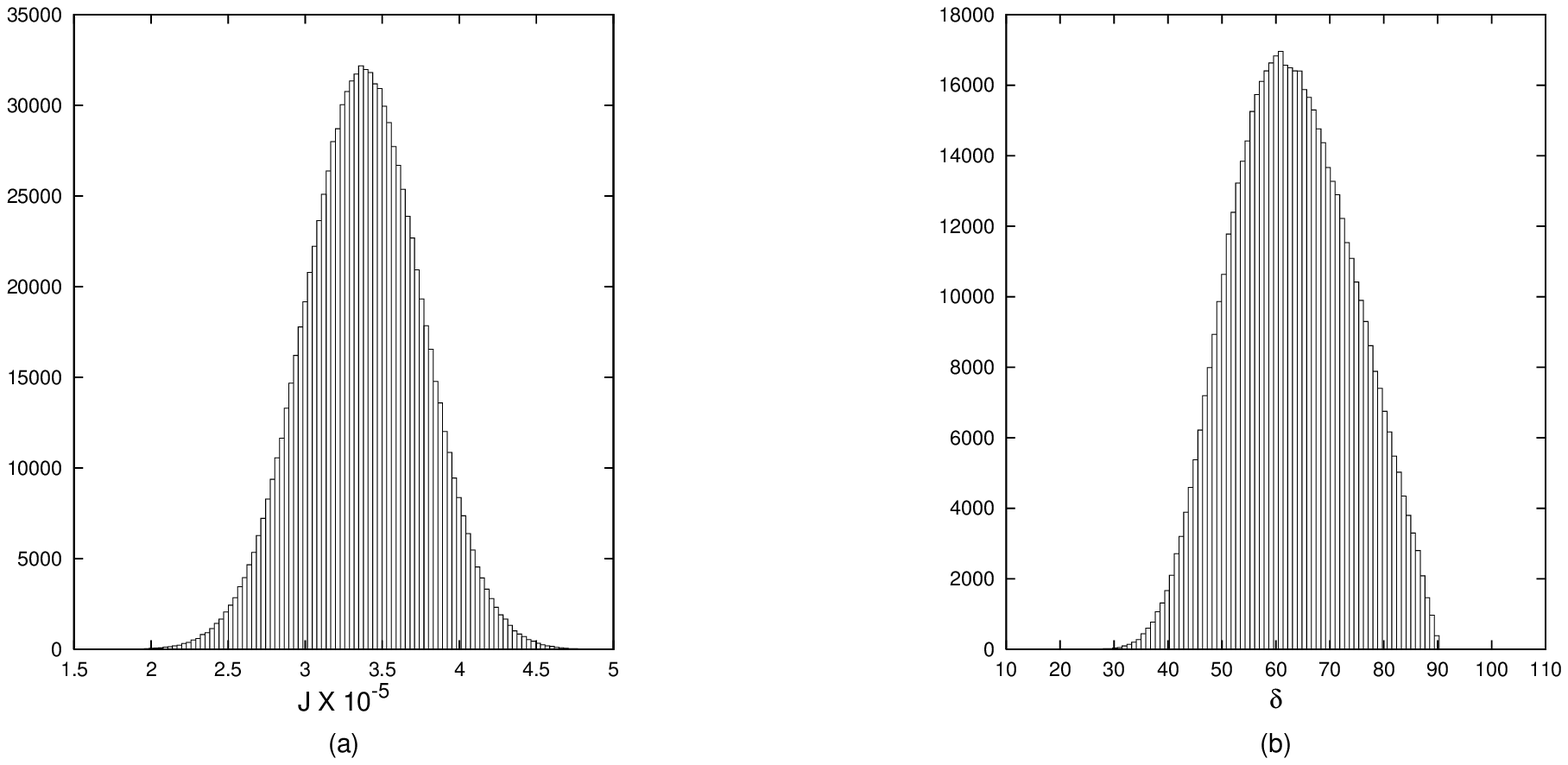}}
\vspace{0.08in}
   \caption{Histogram of $J$ and $\delta$ for `$db$'
triangle in the case of quarks}
  \label{jdelqua}
  \end{figure}

Making use of the above matrix, one can construct the usually
considered `$db$' unitarity triangle in the quark sector,
expressed through the relation
  \be V_{ud} {V_{ub}}^* + V_{cd} {V_{cb}}^* + V_{td}
{V_{tb}}^* =0\,.\label{db} \ee Further, assuming the existence of
CP violation, the Jarlskog's rephasing invariant parameter $J$,
equal to twice the area of the unitarity triangle, can then be
easily constructed using the magnitudes of the elements of the CKM
matrix. Consequently, in the PDG representation, the CP violating
phase $\delta$ can also be evaluated by using the relation between
the parameter $J$ and phase $\delta$, e.g., $
J=s_{12}s_{23}s_{13}c_{12}c_{23}c_{13}^2 \,{\rm sin}\,\delta. $
Following the procedure underlined in \cite{monpro}, we have
obtained a histogram of $J$, shown in figure \ref{jdelqua}a, from
which one can find
\be
J= (3.36 \pm 0.38) \times 10^{-5},\ee the corresponding histogram
of $\delta$, shown in figure \ref{jdelqua}b, yields \be
\delta=62.60^{\rm o} \pm 10.98^{\rm o}. \label{delq} \ee
Interestingly, we find that the above mentioned $J$ and $\delta$
values are compatible with those given by PDG 2010 \cite{pdgnew}.

\section{Lepton mixing case \label{lepton}}
The above discussion as well as the present information regarding
the neutrino mixing angle $s_{13}$ immediately provide a clue for
exploring the possibility of existence of CP violation in the
leptonic sector. We begin with the neutrino mixing matrix
\cite{pmns1,pmns2}, \be \left( \ba{c} \nu_e \\ \nu_{\mu} \\
\nu_{\tau} \ea \right)
  = \left( \ba{ccc} U_{e1} & U_{e2} & U_{e3} \\ U_{\mu 1} & U_{\mu 2} &
  U_{\mu 3} \\ U_{\tau 1} & U_{\tau 2} & U_{\tau 3} \ea \right)
 \left( \ba {c} \nu_1\\ \nu_2 \\ \nu_3 \ea \right),  \label{nm}  \ee
where $ \nu_{e}$, $ \nu_{\mu}$, $ \nu_{\tau}$ are the flavor
eigenstates and $ \nu_1$, $ \nu_2$, $ \nu_3$ are the mass
eigenstates. Following PDG representation, involving three angles
and the Dirac-like CP violating phase $\delta_l$ as well as the
two Majorana phases $\alpha_1$, $\alpha_2$, the PMNS matrix $U$
can be written as \beqn U ={\left( \ba{ccl} c_{12} c_{13} & s_{12}
c_{13} & s_{13}e^{-i \delta_l} \\ - s_{12} c_{23} - c_{12} s_{23}
s_{13} e^{i \delta_l} & c_{12} c_{23} - s_{12} s_{23} s_{13} e^{i
\delta_l} & s_{23} c_{13}
\\ s_{12} s_{23} - c_{12} c_{23} s_{13} e^{i \delta_l} & - c_{12}
s_{23} - s_{12} c_{23} s_{13} e^{i \delta_l} & c_{23} c_{13} \ea
\right)} \left( \ba{ccc} e^{i \alpha_1/2} & 0 & 0 \\ 0 &e^{i
\alpha_2/2} & 0 \\ 0 & 0  & 1 \ea \right). \label{nmm} \eeqn The
Majorana phases $\alpha_1$ and $\alpha_2$ do not play any role in
neutrino oscillations and henceforth would be dropped from the
discussion.

Before coming to the issue of existence of CP violation in the
leptonic sector, we have to first determine the approximate
magnitudes of the elements of the PMNS matrix. To this end,
analogous to the construction of the CKM matrix presented in
Eq.~(\ref{ckmmag}) and using the inputs given in
Eq.~(\ref{s12s23}), we get
 \be U = \left(  \ba{ccc}
  0.8190\pm0.0105 & 0.5516\pm0.0151  &  0.1581\pm0.0221\\
0.4254\pm0.0315  &  0.6317\pm0.0442 & 0.6399\pm0.0610\\
0.3620\pm0.0358  &  0.5376\pm0.0516 & 0.7520\pm0.0519
 \ea \right). \label{r2} \ee
It is interesting to note that the present neutrino mixing matrix
is compatible with those given by
\cite{othersmm1}-\cite{othersmm4}.

\begin{figure}[tbp]
\centerline{\epsfysize=2.6in\epsffile{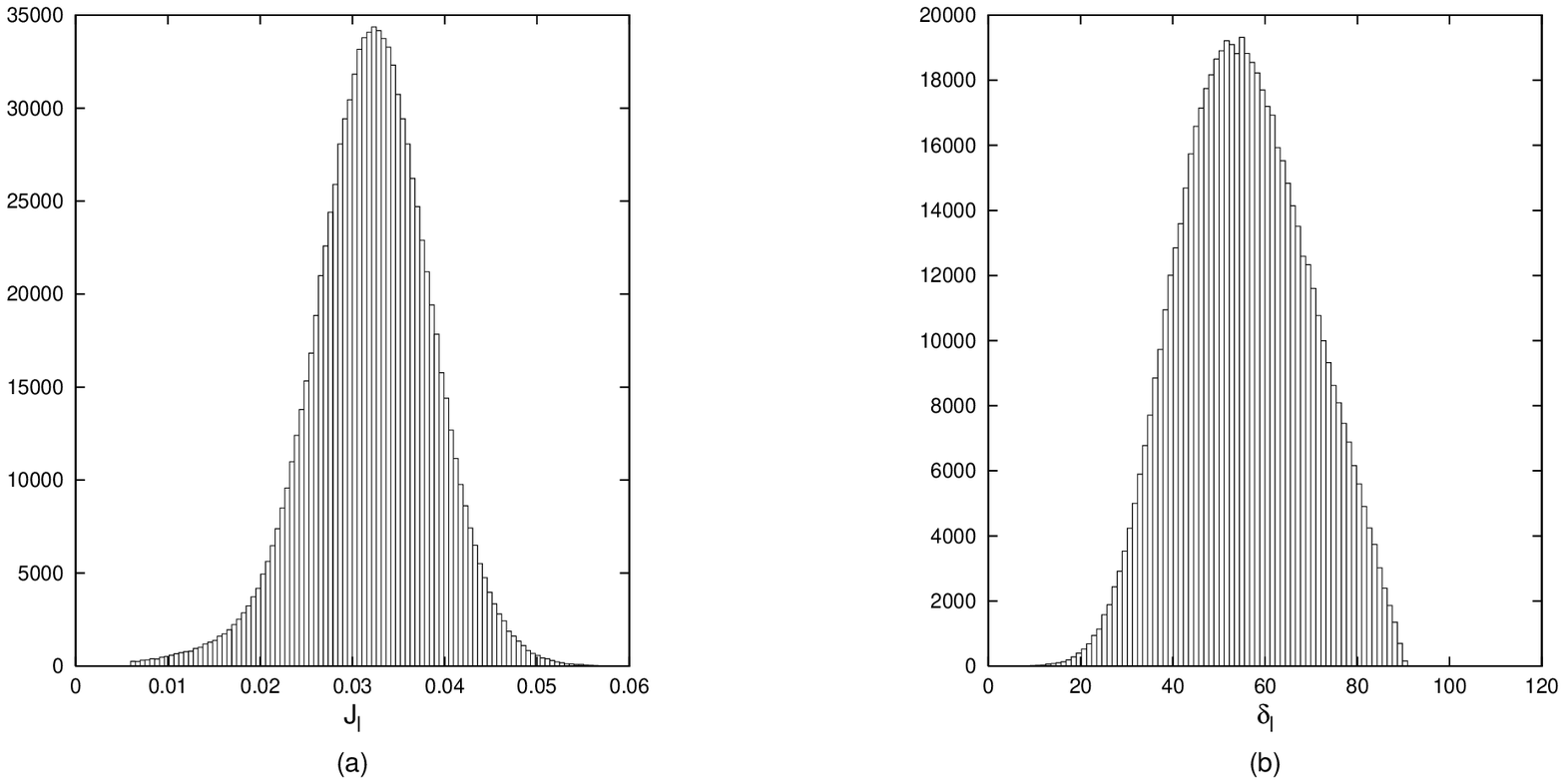}} \vspace{0.08in}
   \caption{Histogram of $J_l$ and $\delta_l$ for `$\nu_1.\nu_3$'
triangle in the case of neutrinos}
  \label{jdeln}
  \end{figure}

Analogous to the `$db$' triangle in the quark sector, we have
considered the `$\nu_1.\nu_3$' unitarity triangle, expressed as
\be U_{e 1}U_{e 3}^{*} + U_{\mu 1}U_{\mu 3}^{*} + U_{\tau
1}U_{\tau 3}^{*}
 = 0 \,. \label{nu1- nu3 } \ee
Following the same procedure as in the quark case, using the
matrix given in Eq.~(\ref{r2}), we obtain the Jarlskog's rephasing
invariant parameter in the leptonic sector $J_l$ as \be J_l=
0.0318 \pm 0.0065, \label{jval}\ee corresponding distribution has
been plotted in figure \ref{jdeln}a. It may be noted that in most
of the recent analyses of neutrino oscillation phenomenology
\cite{jupperbound1}-\cite{jupperbound3}, it is usual to calculate
the upper limit of $J_l$, however, in the present analysis we have
calculated its likely range. Also, we find the corresponding phase
$\delta_l$ from the histogram given in figure \ref{jdeln}b as \be
\delta_l=54.98^{\rm o} \pm 13.81^{\rm o}. \label{deltanu} \ee

Interestingly, this likely value of the phase $\delta_l$ is
largely in agreement with several phenomenological analyses
\cite{smirnov,balaji}. Further, it is interesting to note that the
present analysis carried out purely on phenomenological inputs is
also very much in agreement with several analyses based on
expected outputs from different experimental scenarios
\cite{white}-\cite{sugiyama}. In particular, this value reinforces
the conclusions of Marciano and Parsa \cite{marciano} for the
BNL-Homestake (2540 km) proposal.

\section{Summary and conclusions \label{summ}}
To summarize, in view of the latest T2K and MINOS observations
regarding the mixing angle $s_{13}$ along with the other two well
measured mixing angles $s_{12}$ and $s_{23}$ we have explored the
possibility of existence of CP violation in the leptonic sector.
Taking clues from the construction of the `$db$' unitarity
triangle in the quark sector, we have made an attempt to construct
the `$\nu_1.\nu_3$' leptonic unitarity triangle, suggesting a good
possibility of having non zero CP violation. In particular, we
find likely value of the CP violating phase in the leptonic sector
to be $54.98^{\rm o} \pm 13.81^{\rm o}$.

\section*{Acknowledgments}

The author would like to thank M. M. Gupta for suggesting the
problem and for a careful reading of the manuscript. Thanks are
also due to DST, Government of India for financial support and the
Chairman, Department of Physics for providing facilities to work
in the department.

\end{document}